\documentclass{evn2004}
\usepackage{txfonts}
\begin{document}
\setcounter{page}{325}

   \title{Multi-frequency imaging in VLBI}

   \author{Sergey Likhachev}

   \institute{Astro Space Center of P.N. Lebedev Physical Institute
of Russian Academy of Sciences,Profsojuznaya St. 84/32, 117997,
 Moscow, Russia}

   \abstract{

The new technique, {\it multi-frequency imaging} ( MFI) is developed. 
In VLBI, Multi-Frequency Imaging (MFI) consists of 
multi-frequency synthesis (MFS) and multi-frequency analysis (MFA) 
of the VLBI data obtained from observations on various frequencies. 
A set of linear deconvolution MFI algorithms is described.  The algorithms 
make it possible to obtain high quality images interpolated on any given 
frequency inside any given bandwidth, and to derive reliable estimates of spectral 
indexes for radio sources with continuum spectrum.  Thus MFI approach makes 
it is possible not only to improve the quality and fidelity of the images 
and  also essentially to derive the morphology of the observed radio sources. 
   }

   \maketitle
%

\section{Introduction}

The new technique, {\it multi-frequency imaging} ( MFI), is a 
powerful tool not only for {\it multi-frequency synthesis} (MFS) (Conway et al.\ (\cite{conway})), 
but also for {\it multi-frequency analysis} (MFA) of the VLBI data 
obtained from observations on various frequencies. This tool allows us to 
obtain both {\it high quality images interpolated on a given reference frequency} 
inside of a given bandwidth, as well as reliable estimates of {\it spectral indexes} 
for radio sources with continuum spectrum. This approach is very important 
not only for future Space VLBI missions like Radioastron, but also for ground-based 
VLBI arrays like the EVLA and VLBA.  MFI approach makes it is possible not 
only to improve the quality and fidelity of the images and  
also essentially to derive the morphology of the observed radio sources. 
At the same time, this approach shows that MFI will provide the highest 
angular resolution possible for a given Space VLBI mission. 

\section{Statement of the problem}

Let us consider a linear model for intensity $I_{kpq}=I(x_{p},y_{q},\nu
_{k}) $ of the radio source in a point $(x_{p},y_{q})$ on the observational
frequency $\nu _{k}$:

\[
\begin{array}{l}
I_{kpq}\approx \left( {I_{0}}\right) _{pq}+\left( {I_{1}}\right) _{pq}\beta
_{k}\,+\ldots +\left( {I_{N-1}}\right) _{pq}\cdot (\beta _{k})^{N-1}, \\ 
\quad \\ 
\beta _{k}=\frac{\nu _{k}}{\nu _{0}}-1,\quad k=1,\;2,\;\ldots \;,\;K,
\end{array}
\]

\noindent where $\nu _0 $ is reference frequency corresponding to the
intensity $\left( {I_0 } \right)_{pq} $.

If the intensity $I_{kpq} $ in the point $(x_p ,y_q )$ can be approximated
by power law as

\[
I_{kpq}=\left( {I_{0}}\right) _{pq}\cdot \left( {\frac{\nu _{k}}{\nu _{0}}}%
\right) ^{\alpha _{pq}}, 
\]

then we can present it as

\[
I_{kpq}=\left( {I_{0}}\right) _{pq}e^{\xi _{k}\alpha _{pq}}\approx \left( {
I_{0}}\right) _{pq}\cdot \left( {1+\xi _{k}\alpha _{pq}}\right) 
\]

\noindent where $\xi _k = \ln \;\left( {1 + \beta _k } \right) \approx \beta
_k $,

and thus the spectral indexes $\alpha _{pq}=\alpha \left( {%
x_{p},y_{q}}\right) $ can be obtained as

\[
\left( {I_{1}}\right) _{pq}=\alpha _{pq}\cdot \left( {I_{0}}\right) _{pq} 
\]

Let us consider a target function

\[
\rho =\sum\limits_{k=1}^{K}{\sum\limits_{n=0}^{M-1}{\sum\limits_{m=0}^{M-1}{%
w_{knm}\cdot \left\vert {V_{knm}-\hat{V}_{knm}}\right\vert ^{2}}}},
\]

\noindent where, $w_{knm}=w(u_{n},v_{m},\nu _{k})\geq 0$ are weights, ${
V_{knm}}$, ${\hat{V}_{knm}}$ is a measured and a model visibility function
respectively,
\[
{\hat{V}_{knm}=A}_{k}\cdot \sum\limits_{p,q=o}^{M-1}\left[
\sum\limits_{l=o}^{N-1}\left( \widehat{I_{l}}\right) _{pq}\cdot \beta
_{k}^{l}\right] \cdot \exp \left\{ -2\pi i\cdot \left(
u_{n}x_{p}+v_{m}y_{q}\right) \right\} ,
\]

\noindent where, ${A}_{k}$ is a gain coefficient\bigskip\ \bigskip for k-th
antenna, 
\[
\left( \widehat{I_{l}}\right) _{pq}=\Delta ^{2}\varphi _{pq}\cdot \left(
I_{l}\right) _{pq}\left( 1-x_{p}^{2}-y_{q}^{2}\right) ^{-0.5}, 
\]

$\varphi _{pq}$ is a normalized beam, $\Delta $ is a grid step.

The problem of the optimization can be presented as a solution of \ the
following system of linear equations:

\[
\left( {D_0 } \right)_{pq} = 0,\;\ldots ,\;\left( {D_{N - 1} } \right)_{pq}
= 0 
\]

\noindent for a vector of intensity $\left( \widehat{\mathbf{I}}\right)
_{rt}=\left( {\left( \widehat{\mathbf{I}}{_{0}}\right) _{rt},\;\left( 
\widehat{\mathbf{I}}{_{1}}\right) _{rt},\;\ldots ,\;\left( \widehat{\mathbf{I
}}{_{N-1}}\right) _{rt}}\right) ^{T}$, 

where the $m$-th residual map $\left(
{D_{m}}\right) _{pq}$ can be defined as:

$$
\left( {D_{m}}\right) _{pq} =\sum\limits_{k=1}^{K}{\beta _{k}^{m}\times}
$$

\begin{equation}
\times\left\{ {D_{kpq}}-{\sum\limits_{i=0}^{M-1}{\sum\limits_{l=0}^{M-1}{
B_{k,p-i,q-l}\cdot }}}\sum\limits_{n=0}^{N-1}{\left( \widehat{\mathbf{I}}{
_{n}}\right) _{il}\cdot \beta _{k}^{n}}\right\} ,
\end{equation}

\begin{equation}
m = 0,\;1,\;\ldots ,\;N-1,
\end{equation}

where, 
$$
{D_{kpq}=}\sum\limits_{n,m=o}^{M-1}w_{knm}\cdot V_{knm}\cdot \exp \left\{ 2\pi i\left( u_{n}x_{p}+v_{m}y_{q}\right) \right\}
$$
is a k-th
"dirty" map at the point $\left( x_{p},y_{q}\right) $,

$$
B_{k,p-i,q-l}=\sum\limits_{n,m=o}^{M-1}w_{knm}\times
$$
$$
\times\exp \left\{ 2\pi i\left[
u_{n}\left( x_{p}-x_{i}\right) +v_{m}\left( y_{q}-y_{l}\right) \right]
\right\} 
$$
is a k-th "dirty" beam at the point $\left(
x_{p}-x_{i},y_{q}-y_{l}\right) $.


\section{Solution of the problem}

The solution of the problem can be presented as an iterative procedure for 
a vector \bigskip $\left( \widehat{
\mathbf{I}}\right) _{pq}$:
\[
\left( \widehat{\mathbf{I}}\right) _{pq}^{(s)}=\left( \widehat{\mathbf{I}}
\right) _{pq}^{(s-1)}+\gamma \mathbf{E}^{-1}\cdot \left( \mathbf{D}\right)
_{pq}^{(s-1)},
\]

and the residual maps $\left( \mathbf{D}\right) _{rt}=\left\{ {\left( {D_{0}}
\right) _{rt},\left( {D_{1}}\right) _{rt},...,\left( {D_{N-1}}\right) _{rt}}
\right\} ^{T}$:
\[
\left( \mathbf{D}\right) _{rt}^{(s)}=\left( \mathbf{D}\right) _{rt}^{(s-1)}-
\widehat{\mathbf{B}}_{r-p,t-q}\cdot \left[ \left( \widehat{\mathbf{I}}
\right) _{pq}^{(s)}-\left( \widehat{\mathbf{I}}\right) _{pq}^{(s-1)}\right]
.
\]

Here $\mathbf{E=}\left( E_{ij}\right) $ is a positive defined
matrix of maximum values of \ weighted "dirty" beams, $E_{ij}=\left( 
\widehat{B}_{i+j}\right) _{0,0},i,j=0,...,N-1$; $\gamma $ is a loop gain.
The process of the iteration can be completed if $\varepsilon ^{\left(
s-1\right) }<\varepsilon $, where $\varepsilon $ is a given accuracy.
Otherwise it is necessary to suppose $s=s+1$ and to calculate the next 
$\ \varepsilon ^{\left( s\right) }.$Conditions of the convergence of the
algorithm above is $0<\gamma <2$, $1\leq N\leq K$.

The developed algorithm is nothing other than the \textit{
multi-frequency linear deconvolution (multi-frequency CLEAN)}, itself. This procedure 
described in more detail by Likhachev et al.\ (\cite{likhachev}). 
Notice that the developed algorithm allows to synthesize and analyze of 
high-quality VLBI images directly from the visibility data measured observed on a few frequencies, 
without analyses of the images of the source itself. In case of multi-frequency linear
deconvolution, it is possible to synthesize an image of a radio source at any
intermediate frequency \textit{inside} any given frequency band. Thus, 
\textit{spectral interpolation} of the image is feasible. 
This part of the algorithm is carry out the \textit{synthesis}
of the image itself. However, the algorithm also makes it possible to obtain an 
estimate of the \textit{spectral index} for a given radio source, i.e., it
implements the \textit{\ analysis }of the image. It is clear
that multi-frequency imaging (MFI) will provide the highest angular
resolution possible for any VLBI project due to its improved $\left(
u,v\right) $-coverage.

\section{Implementation of the linear deconvolution algorithm}

The algorithm described above was implemented in the software,
\textbf{\textit{Astro Space Locator (ASL) for Windows}} ({\tt http://platon.asc.rssi.ru/dpd/asl/asl.html}).
It was developed by the Laboratory for Mathematical Methods of the Astro Space Center
(Likhachev et al.\ (\cite{likhachev})).

\end{document}